\documentclass[notitlepage,twocolumn,prl,superscriptaddress]{revtex4-1}
\usepackage{amsmath}
\usepackage{changes}
\usepackage{graphicx,placeins}
\usepackage[space]{grffile} 
\usepackage{gensymb}
\usepackage{bm}
\usepackage{amssymb,amsfonts,amsmath,amsbsy,bm,t1enc,latexsym}
\usepackage{soul}
\usepackage{amssymb} 
\usepackage{amsmath} 
\usepackage{amsfonts}
\usepackage{amsmath}
\usepackage{amssymb}
\usepackage{graphicx}
\RequirePackage{lineno}
\usepackage{siunitx}
\usepackage{hyperref}       
\hypersetup{
    colorlinks=true,
    linkcolor=blue,
    filecolor=magenta,      
    urlcolor=cyan,
}

\usepackage[utf8]{inputenc} 
\begin{document}

\title{Protected generation of dissipative Kerr solitons in supermodes of coupled optical microresonators}


\author{A.~Tikan}
\email{alexey.tikan@epfl.ch}
\affiliation{Institute of Physics, Swiss Federal Institute of Technology Lausanne (EPFL), CH-1015 Lausanne, Switzerland}

\author{A.~Tusnin}
\affiliation{Institute of Physics, Swiss Federal Institute of Technology Lausanne (EPFL), CH-1015 Lausanne, Switzerland}

\author{J.~Riemensberger}
\affiliation{Institute of Physics, Swiss Federal Institute of Technology Lausanne (EPFL), CH-1015 Lausanne, Switzerland}

\author{M.~Churaev}
\affiliation{Institute of Physics, Swiss Federal Institute of Technology Lausanne (EPFL), CH-1015 Lausanne, Switzerland}

\author{X.~Ji}
\affiliation{Institute of Physics, Swiss Federal Institute of Technology Lausanne (EPFL), CH-1015 Lausanne, Switzerland}

\author{K.~Komagata}
\affiliation{Institute of Physics, Swiss Federal Institute of Technology Lausanne (EPFL), CH-1015 Lausanne, Switzerland}
\affiliation{Present address: Laboratoire Temps-Fréquence, Avenue de Bellevaux 51, 2000 Neuchâtel, Switzerland}

\author{R.~N.~Wang}
\affiliation{Institute of Physics, Swiss Federal Institute of Technology Lausanne (EPFL), CH-1015 Lausanne, Switzerland}

\author{J.~Liu}
\affiliation{Institute of Physics, Swiss Federal Institute of Technology Lausanne (EPFL), CH-1015 Lausanne, Switzerland}

\author{T.J.~Kippenberg}
\email{tobias.kippenberg@epfl.ch}
\affiliation{Institute of Physics, Swiss Federal Institute of Technology Lausanne (EPFL), CH-1015 Lausanne, Switzerland}

\date{\today}

\begin{abstract}
The driven-dissipative photonic dimer comprised of two evanescently coupled high-Q microresonators is a fundamental element of multimode soliton lattices. It has demonstrated a variety of emergent nonlinear phenomena including supermode soliton generation, symmetry breaking, and soliton hopping. In this article, we present another aspect of dissipative soliton generation in coupled resonators, revealing the advantages of this system over conventional single resonator platforms. Namely, we show that the accessibility of solitons drastically varies for symmetric and antisymmetric supermode families of the dimer. Linear measurements reveal that the coupling between transverse modes, which gives rise to avoided mode crossings, can be almost completely suppressed. We explain the origin of this phenomenon and show its crucial influence on the dissipative Kerr soliton formation process in lattices of coupled high-Q resonators of any type. Choosing a particular example of the topological Su–Schrieffer–Heeger model, we demonstrate how the edge state can be protected from the interaction with higher-order modes, allowing for the formation of topological Kerr solitons.
\end{abstract}

\maketitle
\section{Introduction}

The analogy between molecules and coupled resonator systems has been discussed in various studies~\cite{bayer1998optical,Boriskina2010Photonic,Zhang2019Electronically}. Similar to the molecular energy surface non-crossings (first pointed out by von Neumann and Wigner~\cite{von1993verhalten} in the early years of molecular quantum mechanics), different eigenmode families of an optical resonator experience avoided mode crossings (AMXs) leading to distortions of initially smooth (i.e. unperturbed) dispersion profile~\cite{Carmon2008Static,savchenkov2012kerr,liu2014investigation,Herr2014Mode}. The dispersion profile is important in a plurality of resonant nonlinear wave-mixing schemes and especially in the context of dissipative Kerr soliton (DKS) generation in resonators possessing $\chi^{(3)}$ nonlinear susceptibility~\cite{Herr2014Mode}.

DKSs are localized stable structures found in driven-dissipative nonlinear resonators~\cite{grelu2012dissipative,Herr2014Temporal}. The shape of DKS is given by the balance between dispersion and Kerr nonlinearity, while its amplitude is fixed due to the balance between losses and parametric gain~\cite{wabnitz1993suppression,barashenkov1996existence,leo2010temporal}. The DKS generation in single optical microresonators~\cite{Herr2014Temporal} that can be integrated on a chip~\cite{brasch2016photonic}, has triggered the development of compact broadband frequency combs for various applications~\cite{suh2016dual,marin2017microresonator,marin2017microresonator,Shen2020Integrated,Riemensberger2020Massively,Kippenberg2018Dissipative}. Even though AMX has been successfully employed for triggering nonlinear dynamics in normal dispersion resonators~\cite{xue2015mode,xue2015normal,kim2019turn,helgason2021dissipative}, to control disorder~\cite{karpov2019dynamics}, and to achieve a quiet point~\cite{yi2017single}, it remains an undesirable effect which disrupts the DKS formation process~\cite{Herr2014Mode} and induces instabilities~\cite{guo2017intermode}.

\begin{figure}
    \centering
    \includegraphics[width=0.8\columnwidth]{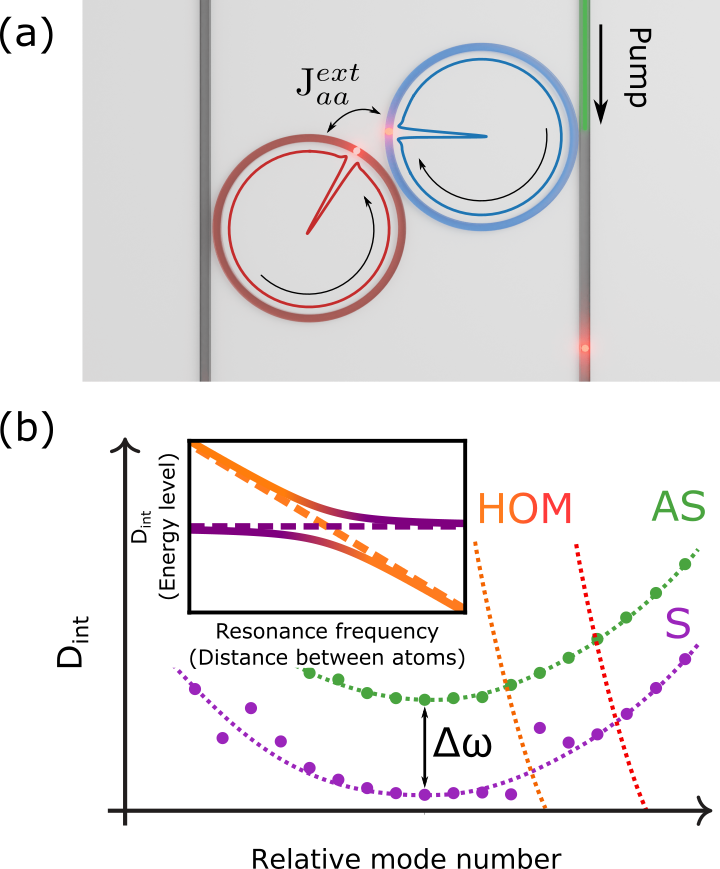}
     \caption{\textbf{Mode interaction in a photonic dimer.} (a) Schematic representation of the driven-dissipative photonic dimer with inter-resonator coupling rate J$^{\mathrm{ext}}_{aa}$ and generated supermode DKS. (b) Schematic dispersion profile of the photonic dimer, showing anomalous modal crossing. HOM stands for higher-order modes, AS — for antisymmetric mode, and S — for the symmetric modes. Inset shows avoided (solid line) and conical (dashed line) crossings in resonators and molecular systems. $D_\mathrm{int}$ stands for the integrated microresonator dispersion. }
    \label{fig:0}
\end{figure}


Recently, the possibility of the DKS generation in a photonic dimer (i.e. a pair of strongly-coupled resonators as shown in Fig.\ref{fig:0}a) has been investigated both experimentally and numerically~\cite{tikan2020emergent,Komagata2021Dissipative}. DKSs can be generated in one of the hybridized dimer supermodes called antisymmetric (AS) and symmetric (S) (see Fig.~\ref{fig:0}b). 
Theoretical studies have predicted a variety of emergent nonlinear phenomena including periodic appearance of commensurate and incommensurate dispersive waves, symmetry breaking, and soliton hopping.
While AS supermode DKS generation has been readily achieved in experiments, some of the predicted effects (such as soliton hopping) were not observed experimentally due to an unexpectedly enhanced interaction of the S supermode family with higher-order modes (HOMs). The influence of AMX on the S supermodes strongly disrupts DKS generation and completely suppresses it after a certain power level. Contrary, the interaction of AS modes with HOMs (see Fig.~\ref{fig:0}b) can be completely eliminated. 

In this work, we investigate the effect of protection against AMXs that are the primary obstacle for the experimental realization of recently introduced nested solitons in high-Q coupled resonator optical waveguides~\cite{tusnin2021coherent} and topological lattices~\cite{mittal2021topological}. We provide an experimental study of the DKS generation in different supermodes of a photonic molecule realized with integrated Si$_3$N$_4$ microresonators~\cite{liu2020photonic}, which confirms our observation. Investigating building blocks of the soliton lattices (broadly hybridized coupled trimer and plaquette of resonators), we prove the universal nature of this effect and demonstrate the DKS generation in the protected mode of the degenerate resonator plaquette. We propose a general model explaining the effect of protection and apply it to the topological Su–Schrieffer–Heeger (SSH) arrangement, demonstrating the failure of the topological protection. Furthermore, we provide a recipe for harnessing the effect for on-demand protection of the dispersion profile, which is essential for the experimental generation of edge state soliton frequency combs.

\begin{figure*}
    \centering
    \includegraphics[width=0.99\textwidth]{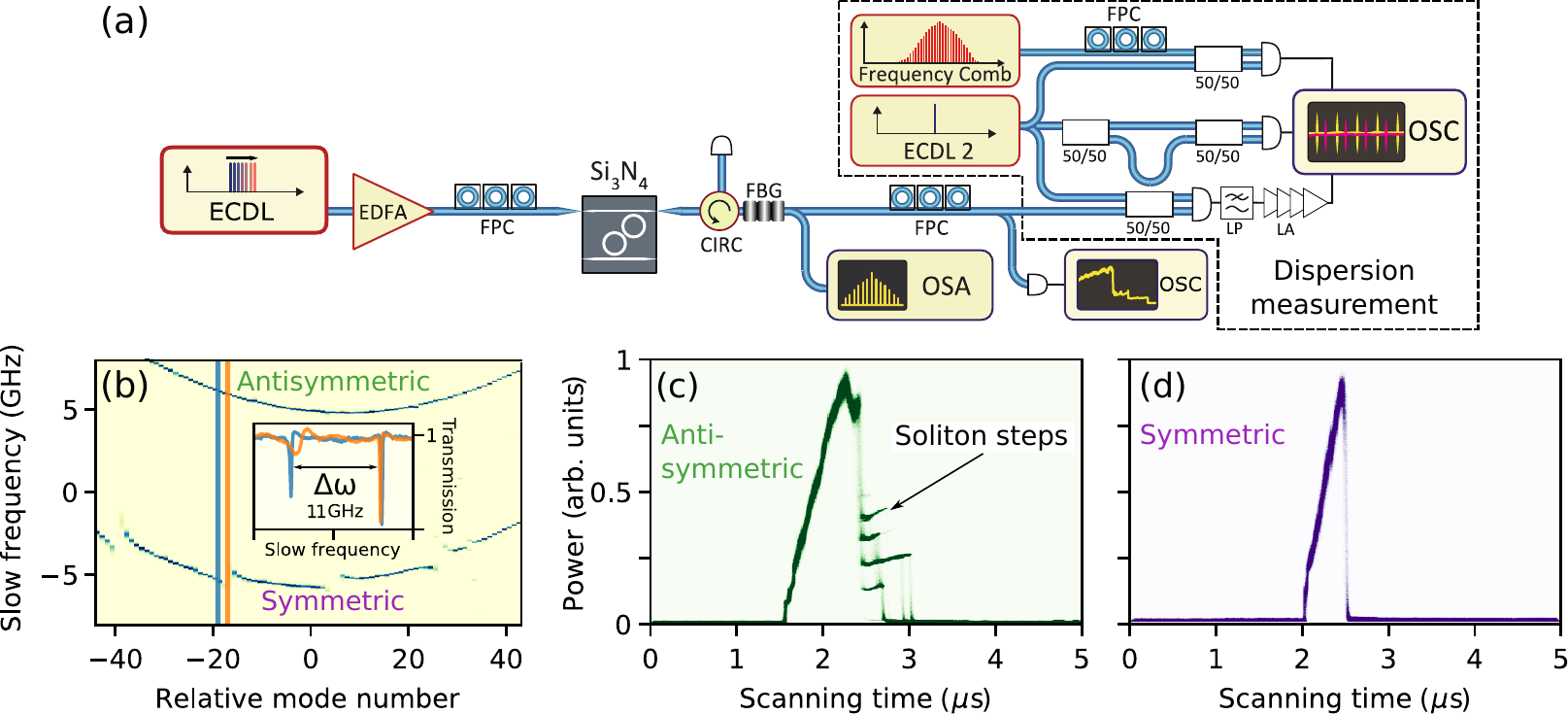}
    \caption{\textbf{Effect of protection against mode crossings in the photonic dimer and supermode DKS generation.} \textbf{(a)} Schematic representation of the experimental setup. Abbreviations: External cavity diode laser (ECDL); erbium doped fiber amplifier (EDFA); fiber polarization controller (FPC); optical circulator (CIRC); fiber Bragg grating (FBG); optical spectrum analyzer (OSA); low-pass filter (LP); logarithmic amplifier (LA); sampling oscilloscope (OSC). ECDL is coupled into and out of the microresonator chip via lensed fibers. (b) at modes -17 and -19 corresponding to AMX and its close vicinity, respectively. \textbf{(c,d)} Experimental recording of generated light in AS and S modes of the photonic dimer, respectively. Each plot contains 600 superimposed oscilloscope traces.
    }
    \label{fig:1}
\end{figure*}

\section{Results}
\subsection{Photonic dimer}
Experimental evidences of the protection effect are obtained with strongly coupled microrings having $\approx$200 GHz free spectral range and loaded Q-factor of the order of 2 millions, realized on the Si$_3$N$_4$ platform using the photonic Damascene reflow process~\cite{Pfeiffer2018Ultra}. The intrinsic loss rate of the dimer is 50 MHz, and both resonators are interfaced with bus waveguides having external coupling rates of 100 MHz. Employing frequency comb calibrated diode laser spectroscopy (see Fig.~\ref{fig:1}a as well ~\cite{del2009frequency}), we retrieve first the dispersion profile of the photonic dimer (Fig.~\ref{fig:1}b). It is represented in terms of slow frequency, which has been defined as integrated dispersion (D$_\mathrm{int}$) normalized by $2\pi$ with an arbitrary shift. The integrated dispersion is defined as $D_\mathrm{int}(\mu) = \omega_\mu - (\omega_0 + D_1\mu)$, where $\omega_\mu$and $\mu$ stand for the resonance frequency and the corresponding index, $D_1/(2\pi)$ is the free spectral range of the resonator at frequency $\omega_{0}$. The dispersion profile reveals two dimer supermode families. The fundamental S supermode family is strongly affected by AMXs while the AS supermode dispersion profile is almost unperturbed. The inset of Fig.~\ref{fig:1}b shows a cross-section of the plot at the mode crossed by a HOM (orange line) and before it (blue line).
HOMs and, therefore, AMXs are likely to be present in soliton-generating microresonators because the waveguide constituting the microresonator is usually chosen to be multimode in order to guarantee low propagation losses (hence high Q factor) by reducing the influence of the fabrication-induced surface roughness~\cite{Ji2021Exploiting}. 

We also study the influence of AMXs on the generation of supermode DKSs. Fig.~\ref{fig:1}c,d show a superposition of 600 transmission traces obtained by sweeping the pump laser frequency over the AS and S resonances at 1554 nm with an optical power in the waveguide of 43 mW. We use a conventional CW pumping scheme combined with fast single side-band tuning to eliminate thermal heating and resonance shifts~\cite{stone2018thermal}. Generated light profiles for the AS mode, detected with a photodiode after filtering out the pump comb line, systematically show the presence of characteristic steps signifying the stable access to the solitonic state~\cite{Herr2014Temporal,guo2017universal}. Contrary, S supermodes exhibit no solitonic feature at the equivalent pump power. At lower input powers, soliton generation can  be observed in the S mode family, however, it depends on the particular distribution of AMXs on the dispersion profile.
Indeed, as pointed out in~\cite{bao2017spatial} presence of the AMX leads to intense generation of dispersive waves which perturb the solitonic state. Each soliton acts as a source of dispersive waves and, therefore, the number of solitons is naturally reduced until the perturbation becomes sufficiently weak to maintain the state. The strength of the perturbation depends on the position of AMX since the power spectral density of the soliton and, hence, the optical power transferred to the HOM decays exponentially from the pumped mode.

\subsection{Photonic trimer and plaquette}
\begin{figure*}
    \centering
    \includegraphics[width=0.99\textwidth]{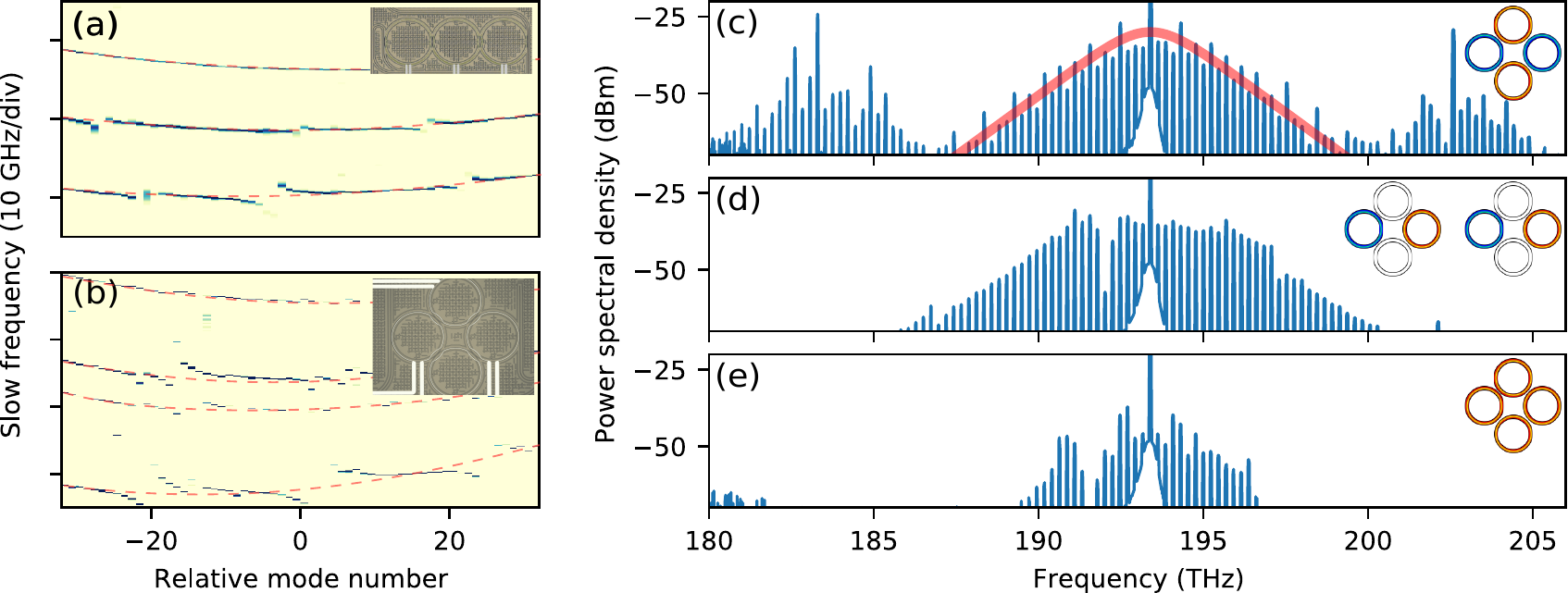}
    \caption{\textbf{Effect of protection in photonic trimer and plaquette.} \textbf{(a)} Linear dispersion measurement of a chain of three coupled resonators (photonic trimer). \textbf{(b)} The same for a square lattice (plaquette). Insets show microscope images of the Si$_3$N$_4$ microresonators of $\approx$200 GHz free spectral range and imprinted system of heaters. \textbf{(d-e)} Optical spectra obtained by investigating the square lattice tuned into the degenerate state similar to the trimer configuration. Spectral corresponds to top, middle and bottom resonances of the effective trimer, respectively. The red line shows a fitting with the hyperbolic secant profile. Insets show a schematic representation of the supermode distribution.}
    \label{fig:1_2}
\end{figure*}

A similar effect is observed in the trimer configuration. Linear dispersion measurements (see Fig.~\ref{fig:1_2}a) reveal that the protection effect is the strongest for the trimer supermode with the highest relative frequency and gradually decreases for lower frequency states on the integrated dispersion profile. We also investigate a more complex resonator arrangement representing a fundamental element of the square lattice - a plaquette. Fig.~\ref{fig:1_2}b shows the corresponding dispersion profile. In the ideal case, two central mode families are degenerate. However, due to the presence of the finite inter-resonator detuning $\delta$, the degeneracy is lifted and we observe all four mode families. Imprinting metallic heater on top of the resonators, we can control the inter-resonator detuning and, therefore, can tune the plaquette system to the degenerate state. We choose to work with this system as it represents a more general case of coupled resonators, including the trimer to the degenerate case. Fig.~\ref{fig:1_2}c-e corresponding to the upper, middle (2x degenerate) and lower frequency resonance, show nonlinear probing of the plaquette structure. As follows from the figure, DKS in such configuration can be generated only in the upper resonance, which is expected to be protected from interaction with HOM. 

\begin{figure*}
    \centering
    \includegraphics[width=0.99\textwidth]{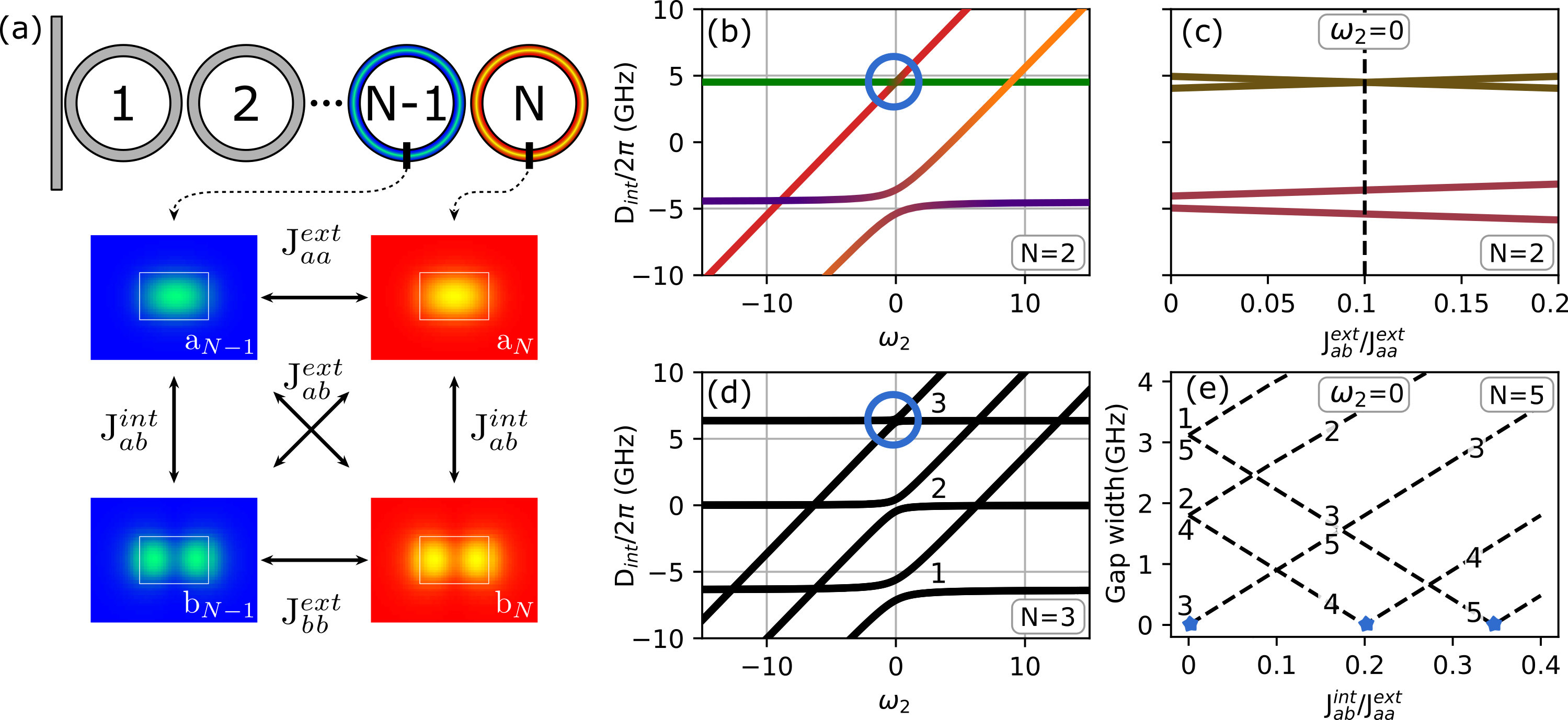}
    \caption{\textbf{Effect of protection in coupled resonators.} \textbf{(a)} Schematic representation of coupled resonators chain and description of the coefficients of the coupling matrix used in Eq.~\ref{eq1}. \textbf{(b)} Protection in the photonic dimer. Protected state is highlighted by the blue circle. Green line corresponds to AS fundamental mode, violet to S fundamental mode, red — AS higher-order mode, and orange — S higher-order mode. Parameters are chosen to be close to the experimental ones: $J_{aa}^{\mathrm{ext}}/2\pi$ = 4.5 GHz, $J_{ab}^{\mathrm{int}}$ = $J_{ab}^{\mathrm{ext}}$ = 0.1$J_{aa}^{\mathrm{ext}}$ \textbf{(c)} The splitting between hybridized supermodes with coinciding central frequencies. Dashed black line shows the parameters corresponding to plot (b). \textbf{(d}) Protection in the trimer configuration. \textbf{(e)} The gap distance between split resonances as a function of $J_{ab}^{\mathrm{int}}$/$J_{aa}^{\mathrm{ext}}$ for a chain of 5 coupled resonators. Lines are numbered according to the relative frequency of the modes, as shown in plot (d). Blue stars depict the protected states.}
    \label{fig:2}
\end{figure*}

\subsection{Model of mode crossing suppression}
In order to shed light on the protection phenomenon, we derive from Maxwell's equations a Hermitian model of four coupled modes interaction (see Supplementary Materials). We consider two fundamental $a_{1(2)}$ and two transverse HOMs $b_{1(2)}$ of both resonators constituting the dimer. The coupled mode equations can be written as follows~\cite{haus1991coupled}:

\begin{equation}\label{eq1}
i\frac{dU}{dt}=
-\left(
\begin{array}{cccc}
-\omega_1 & J_{aa}^{\mathrm{ext}} & J_{ab}^{\mathrm{int}} & J_{ab}^\mathrm{ext} \\
J_{aa}^{\mathrm{ext}} & -\omega_1 & J_{ab}^{\mathrm{ext}} & J_{ab}^{\mathrm{int}} \\
J_{ba}^{\mathrm{int}} & J_{ba}^{\mathrm{ext}} & -\omega_2 & J_{bb}^{\mathrm{ext}} \\
J_{ba}^{\mathrm{ext}} & J_{ba}^{\mathrm{int}} & J_{bb}^{\mathrm{ext}} & -\omega_2
\end{array}
\right)U,
\end{equation}
where U = $(a_1,a_2, b_1, b_2)^\intercal$. Eigenvalues of the coupling matrix can be found analytically. Assuming that the coupling matrix is symmetric, we obtain:
\begin{equation} \label{eq2}
\begin{split}
& \lambda_{1,2}(\mathrm{as}) = \\ 
& \frac{1}{2} \left(2 J_{aa}^{\mathrm{ext}}\pm\sqrt{4(J_{ab}^{\mathrm{int}}-J_{ab}^{\mathrm{ext}})^2+\left(\omega_1-\omega_2\right)^2}+\omega_1+\omega_2\right)\\
& \lambda_{3,4}(\mathrm{s}) = \\ 
& \frac{1}{2} \left(-2 J_{aa}^{\mathrm{ext}}\pm \sqrt{4(J_{ab}^{\mathrm{int}}+J_{ab}^{\mathrm{ext}})^2+\left(\omega _1-\omega_2\right)^2}+\omega_1+\omega _2\right).
\end{split}
\end{equation}
The notation for coupling coefficients is described in Fig.~\ref{fig:2}a. $J_{aa}^{\mathrm{ext}}$ corresponds to the coupling between fundamental modes of the nearest resonators, $J_{ab}^{\mathrm{ext}}$ — to the coupling between fundamental mode of one resonator and HOM of the neighbor, and  $J_{ab}^{\mathrm{int}}$ is the coupling between fundamental and HOM within the same resonator. The coupling strength between two HOMs is set to $J_{aa}^{\mathrm{ext}}$ since it does not qualitatively change the result. The difference between $J_{aa}^{\mathrm{ext}}$ and  $J_{bb}^{\mathrm{ext}}$ leads to a shift of the hybridization area along the direction of the HOM. As an example of the resonator HOM, we show TE$_{10}$.

Thus, we find two pairs of eigenvalues that represent the mode interaction. The first couple $\lambda_{1,2}$ corresponds to the AS supermodes while $\lambda_{3,4}$ to S one. The expression under the square root in the first couple of eigenvalues contains the term $(J_{ab}^{\mathrm{int}} - J_{ab}^{\mathrm{ext}})^2$. Therefore, in the case when $J_{ab}^{\mathrm{int}}$ and $J_{ab}^{\mathrm{ext}}$ are of the same order, the influence of the AMX is reduced. However, in the second couple eigenvalues, the effect of AMX is increased in comparison to the conventional hybridization in the single resonator case. 

Indeed, numerical simulations of the coupling region for parameters close to experimental ones (see Supplementary Materials) demonstrated that the ratio between $J_{ab}^{\mathrm{ext}}$ and $J_{ab}^{\mathrm{int}}$ tends to one. The coupling sections to bus and drop waveguides will contribute to the coefficient $J_{ab}^{\mathrm{int}}$ as well, however, this contribution is found to be one order of magnitude smaller which is consistent with the experimentally observed strong protection of AS supermode parabola (Fig.~\ref{fig:1}b).

The eigenvalue system Eq.~\ref{eq2} as a function of the central frequency of the HOM $\omega_2$ with the ratio $J_{ab}^{\mathrm{int}}/J_{ab}^{\mathrm{ext}} = 1$ is depicted in Fig.~\ref{fig:2}b. Fig.~\ref{fig:2}c shows the dependence of the hybridized modes position for $\omega_1=\omega_2$ (at the center of Fig.~\ref{fig:2}b) as a function of $J_{ab}^{\mathrm{ext}}$/$J_{aa}^{\mathrm{ext}}$. Black dashed line corresponds to the conditions considered in Fig.~\ref{fig:2}b. As predicted from Eq.~\ref{eq2}, when $J_{ab}^{\mathrm{int}}$ and $J_{ab}^{\mathrm{ext}}$ coincide exactly, the gap distance tends to zero. 

The structure of the coupling matrix in Eq.~\ref{eq1} is notably similar to the Hamiltonian discussed in~\cite{hatton1976non}, which underpins the profound nature of the analogy with molecular systems. Similar effects, known as conical or diabolical crossings~\cite{yarkony1996diabolical} in this community, have been actively investigated at the end of the last century.

This model can be easily extended to the case of arbitrary number of resonators (see Supplementary Materials). An example of the mode hybridization for the trimer configurations is shown in Fig.~\ref{fig:2}d. Influence of the AMX increases with the decreasing relative frequency, as suggested by the experimental data. Numerical analysis of longer chains revealed that the index of the protected mode depends on the values $J_{ab}^{\mathrm{ext}}$ and $J_{ab}^{\mathrm{int}}$ and, therefore, \emph{can be manipulated}. When $J_{ab}^{\mathrm{ext}}$ can be neglected then the effect of AMXs is the same for all the hybridized mode. In the opposite case, when $J_{ab}^{\mathrm{int}} \ll J_{ab}^{\mathrm{ext}}$, the protection falls into the middle mode family and symmetrically decreases towards modes with higher and lower relative frequencies. Therefore, the protection can be moved along the dispersion relation by changing the coupling coefficients ratio. Fig.~\ref{fig:2}e shows the dependence of the AMX-induced gap width as a function of normalized $J_{ab}^{\mathrm{int}}$ coefficient for a five resonator chain. When $J_{ab}^{\mathrm{int}}$ can be neglected, the middle mode (3) becomes protected. With increasing $J_{ab}^{\mathrm{int}}$, the protection moves towards the fourth mode family and subsequently to the fifth one.

\begin{figure}
    \centering
    \includegraphics[width=0.8\columnwidth]{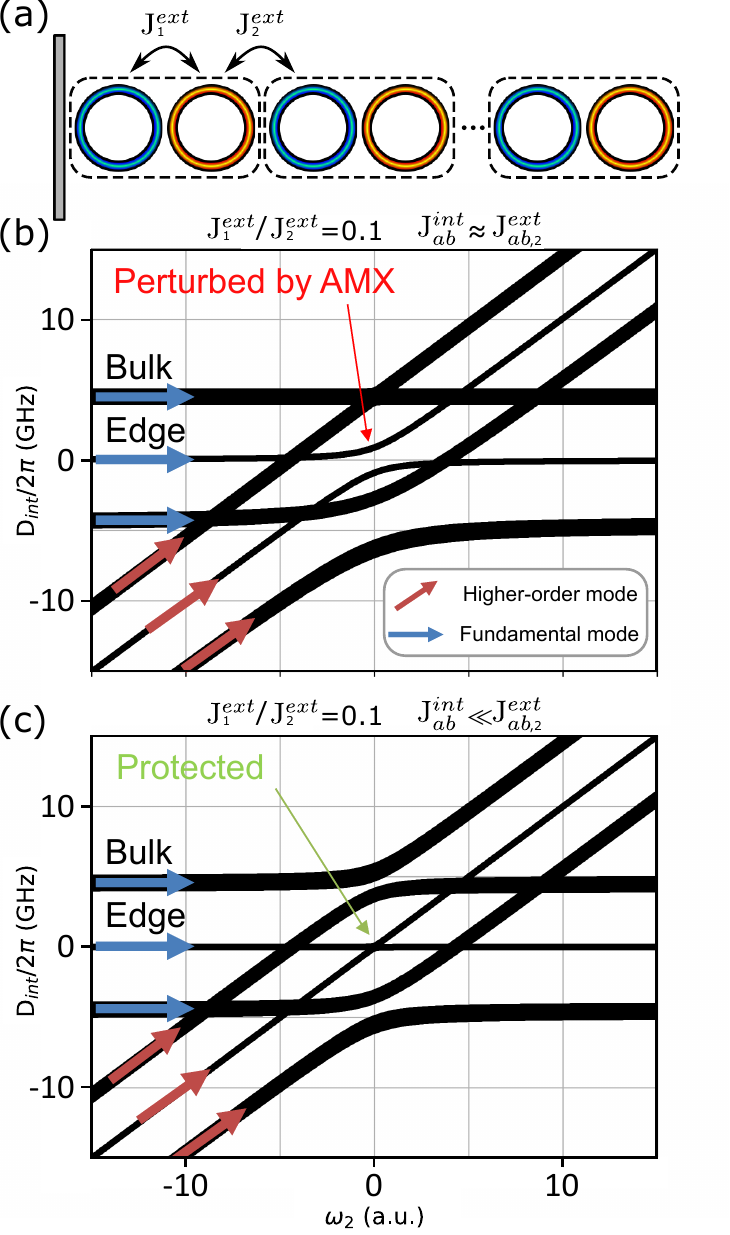}
    \caption{\textbf{Protection against mode crossing of the topological edge state in Su–Schrieffer–Heeger model.} \textbf{(a)} Schematic representation of 10 coupled resonators with alternating coupling, which constitutes the SSH chain. \textbf{(b)} Influence of AMX when $J_{ab}^{\mathrm{int}} \approx J_{ab,2}^{\mathrm{ext}}$. \textbf{(c)} The same configuration with $J_{ab}^{\mathrm{int}} \ll J_{ab,2}^{\mathrm{ext}}$ ($J_{ab}^{\mathrm{int}} = 0.001 J_{ab,2}^{\mathrm{ext}}$) exhibits the protection against AMX.}
    \label{fig:3}
\end{figure}

\subsection{Protection of topological states in the Su–Schrieffer–Heeger model}
In order to highlight the importance of protection for the future development of field of the soliton lattices, we study the effect of AMXs on topologically protected edge states in the SSH model~\cite{su1979solitons} originally proposed for the explanation of mobile neutral defects in polyacetylene. This model is described by the following Hamiltonian: $\hat{H}_{\mathrm{SSH}} = \sum_{n} t_1\hat{c}_n^\dagger \hat{c}_{n-1}+t_2\hat{c}_{n+1}^\dagger\hat{c}_{n} $, where $\hat{c}^{\dagger}_n$ is the creation operator at the n$_{th}$ site and $t_{1,2}$ is the hopping amplitudes. Due to the simplicity of implementation~\cite{dobrykh2018nonlinear}, this model often serves as a primary verification platform for novel nonlinearity-related topological effects~\cite{smirnova2020nonlinear}. It can be realized in our system by varying the inter-resonator coupling coefficients $J_{aa}^{\mathrm{ext}}$ and $J_{ab}^{\mathrm{ext}}$ playing here the role of the hopping amplitudes. Schematic representation of the SSH chain is shown in Fig.~\ref{fig:3}a. The alternating coupling effectively divides the chain into a number of unit cells shown by dashed rectangles. The coupling strength ratio inside a unit cell and between unit cells ($t_{1}/t_{2} =J_{aa,ab,1}^{\mathrm{ext}}/J_{aa,ab,2}^{\mathrm{ext}}$) is chosen to be 0.1 which is sufficient for obtaining a nonzero winding number~\cite{asboth2016short}.

Fig.~\ref{fig:3}b shows the mode hybridization for the SSH model realized with ten resonators. The topologically protected (against the disorder and variation of the coupling coefficients) edge state is in the middle of the gap between two bulk states as shown in Fig.~\ref{fig:3}b. The same is observed for the HOM family. According to the model, a crucial influence of AMX on the topological state is expected, which potentially forbids or drastically obstructs the generation of topological DKSs if the protection is not controlled. Indeed, AMXs in this system, together with inter-resonator detuning in absolute frequency, act as on-side potential breaking the chiral symmetry of the SSH lattice~\cite{Chiu2016Classification} and omitting the inherent topological protection~\cite{st2017lasing}. However, increasing the difference between $J_{ab}^{\mathrm{int}}$ and $J_{ab}^{\mathrm{ext}}$, we observe enhancement of the protection effect on the edge state modes. This can be achieved by carefully designing the coupling section to suppress the internal transverse mode couplings. Choosing the ratio $J_{ab}^{\mathrm{int}} = 0.001 J_{ab,2}^{\mathrm{ext}}$, we observe a complete protection against AMX as shown in Fig.~\ref{fig:3}c.


\section{Conclusion}
Concluding, we introduced the notion of protection against AMX in chains of coupled multimode resonators, which exhibit a remarkable similarity with conical energy level crossings in molecular systems. The crucial influence of this effect on the dispersion profile and, therefore, the supermode DKS generation is demonstrated experimentally. We propose a simple model which fully explains the effect and proposes a way to harness it for stable DKS generation. We highlight its importance by showing that topologically protected edge state in the SSH model can be highly influenced by AMXs and, therefore, it must be carefully taken into the account while designing the experimental platform for the observation of topological solitons.

\section{Acknowledgments}
This publication was supported by Contract 18AC00032 (DRINQS) from the Defense Advanced Research Projects Agency (DARPA), Defense Sciences Office (DSO). This material is based upon work supported by the Air Force Office of Scientific Research under award number FA9550-19-1-0250. This work was further supported by the European Union’s Horizon 2020 Research and Innovation Program under the Marie Skłodowska-Curie grant agreement 812818 (MICROCOMB), and by the Swiss National Science Foundation under grant agreements 192293 and 176563. Si$_3$N$_4$ samples were fabricated and grown in the Center of MicroNanoTechnology (CMi) 
\section{Supplemental document} See Supplement 1 for supporting content.


%

\end{document}


\title{Supplementary Materials: Protected generation of dissipative Kerr solitons in  supermodes of coupled optical microresonators}

\author{A.~Tikan$^{*1}$, A.~Tusnin$^1$, J.~Riemensberger$^1$, M.~Churaev$^1$, X.~Ji$^1$, K.~Komagata$^{1,2}$, R.~N.~Wang$^1$, J.~Liu$^1$, T.J.~Kippenberg$^*$}
\affiliation{Institute of Physics, Swiss Federal Institute of Technology Lausanne (EPFL), CH-1015 Lausanne, Switzerland \\ $^2$Present address: Laboratoire Temps-Fréquence, Avenue de Bellevaux 51, 2000 Neuchâtel, Switzerland}
\date{\today}
\email{alexey.tikan@epfl.ch, tobias.kippenberg@epfl.ch}
\pacs{}

\maketitle

\section{Mode interactions in two evanescently coupled resonators}

We examine here the mode interaction of two ring resonators using a perturbation approach. Taking the solution of the Maxwell's equations for a single resonator as a basis for the field profile in the coupled system, we derive coupling coefficients between different modes as a function of mode overlaps. The obtained expressions can be used for the exact evaluation of the coupling coefficients for whispering gallery mode resonators, since analytical expressions are known in this case~\cite{GorodetskyGeometrical,SturmanVectorial}, while for integrated microresonator dimers they, however, help to qualitatively understand the mode-crossing mechanism.

We start from the scalar wave equation on electric field in the system comprising two identical evanescently coupled optical resonators. The wave equation 
\begin{equation}\label{eq:wave_eq}
    (\Delta + \frac{n_g^2}{c^2}\frac{\partial^2}{\partial t^2})E =0
\end{equation}
governs the electric field in the media with the group index $n_g$. In a single resonator case, rotational symmetry allows one to obtain a set of eigenfrequencies  and eigenfunctions supported by the system. Typically, there are two polarization mode families (TE and TM), and within each polarization mode family there is a set of eigenfunctions (i.e. states) which have different spatial distributions.  In the ideal case, all the eigenfunctions are orthogonal, even if they correspond to degenerate eigenfrequencies. The presence of perturbations which cause the axial symmetry breaking leads to the interaction between the modes breaking, thereby their orthogonality. This interaction manifest itself as avoided mode crossings (AMXs) in the dispersion profile. The AMXs appear at degenerate frequencies, where two different modes have close eigenvalues~\cite{zhu2010chip,Strekalov2016nonlinear}. 

The scenario of mode interaction in the photonic dimer is similar to the conventional single resonator case, but the difference arises from the fact that we investigate two sets of eigenmodes which belong to different rings. In order to obtain the coupling coefficients, we employ the perturbation approach. Starting from independent eigenfunctions for both rings, implying at first the infinite distance between the resonators, we decompose the electric field on a series of eigenfunctions and obtain a system of coupled ordinary differential equations on amplitudes of the modes.

We start with a singe resonator case which has the group index $n_{g_I}$. Using the ansatz of harmonic time dependence $E\to E e^{-i\omega t}$, one obtains the Helmholtz equation 
\begin{equation}\label{eq:helmholtz}
    (\Delta + n_{g_I}^2k_0^2)E = 0,
\end{equation}
where $k_0 = \omega/c$ is the wavenumber. Eq.~(\ref{eq:helmholtz}) defines eigenfrequencies $\omega_\mu^I$ and eigenfunctions $\Psi_\mu^I$ with orthogonality relation:
\begin{equation}\label{eq:orth_rel}
    \int \Psi_\mu^I (\Psi_\nu^I)^* n_{g_I}^2 dV = \delta_{\mu,\nu}.
\end{equation}
Here asterisk stands for complex conjugation. The electric field is $E = A^I_\mu \Psi^I_\mu$, where $A^I_\mu$ is the normalization constant. Note, the same is valid for the second resonator, for which it is sufficient simply to replace $I$ by $II$. Moreover, we consider the resonators to be identical, meaning that $\omega_\nu^I = \omega_\nu^{II}$.

Now, we suppose that the two resonators are placed closed so their eigenfunctions overlap. In order to exploit the eigenfuctions of each ring, we rewrite the group index $n_g$ in the following form:
\begin{equation}
n_g^2=
    \begin{cases}
    n_{g_I}^2 + n_{II}\\
    n_{g_{II}}^2 + n_{I},
    \end{cases}
\end{equation} 
depending on the basis we want to use. We decompose further the electric field in the following way:
\begin{equation}
    E = \sum_{i}A^I_i(t) \Psi_i^I e^{-i \omega_i t} + \sum_{i}A^{II}_i(t) \Psi_i^{II} e^{-i \omega_i t}.
\end{equation}
Substituting this to the Eq.~\ref{eq:wave_eq} and using the slowly varying envelope approximation ($d^2 A_i^I/dt^2 \ll \omega_i d A^I/dt$) we obtain the following equation:
\begin{align}\label{eq:gen_system}
    &\Delta E - \frac{n_g^2}{c^2}\frac{\partial^2 E}{\partial t^2} = \nonumber \\
    &\sum_i\Big(n_{II} \Psi_i^I A_i^I \lambda_i^2  + 2i\lambda_i \dot{A}_i^I \Psi_i^I \frac{(n_{g_I}^2 + n_{II})}{c} \Big)e^{-i\omega_i t}+ \nonumber \\
    &\sum_i\Big(n_{I} \Psi_i^{II} A_i^{II} \lambda_i^2  + 2i\lambda_i \dot{A}_i^{II} \Psi_i^{II} \frac{(n_{g_{II}}^2 + n_{I})}{c} \Big)e^{-i\omega_i t}=0,
\end{align}
where $\lambda_i = \omega_i/c$ and $\dot{A}$ stands for the time derivative of $A$. Now, we multiply this equation by $(\Psi_k^{I(II)})^*$ and integrate it over the whole volume. Using the orthogonality relation~(\ref{eq:orth_rel}), one obtains a system of ordinary differential equations on the mode amplitudes with coupling coefficients proportional to the mode overlap. 

Considering the case of two mode families in both resonators, one can derive the matrix model introduced in the main text. In order to keep the same notations, we denote $A_1^I \equiv a_1$ ,$A_2^I \equiv b_1$, $A_1^{II} \equiv a_2$, $A_2^{II} \equiv b_2$. Taking only leading order coefficients, system~(\ref{eq:gen_system}) takes form
\begin{align}\label{eq:sys_1}
    \begin{cases}
    \dot{a}_1 &= i (J_{a_1 a_1} a_1 + J_{a_1 a_2}a_2 + J_{a_1 b_1} b_1 + J_{a_1 b_2}b_2)\\
    \dot{a}_2 &= i (J_{a_2 a_1} a_1 + J_{a_2 a_2}a_2 + J_{a_2 b_1} b_1 + J_{a_2 b_2}b_2)\\
    \dot{b}_1 &= i (J_{b_1 a_1} a_1 + J_{b_1 a_2}a_2 + J_{b_1 b_1} b_1 + J_{b_1 b_2}b_2)\\
    \dot{b}_2 &= i (J_{b_2 a_1} a_1 + J_{b_2 a_2}a_2 + J_{b_2 b_1} b_1 + J_{b_2 b_2}b_2),\end{cases}
\end{align}
where diagonal terms indicate self-frequency shift due to presence of the coupling sections. They can be expressed through the mode overlap integrals as follows:
\begin{align}\label{eq:self_freq}
    J_{a_1 a_1}& = \frac{\lambda_0 c}{2}\int \Psi_0^{I}\Psi_0^{I^*}n_{II}dV;\, 
    J_{a_2 a_2} = \frac{\lambda_0 c}{2}\int \Psi_0^{II}\Psi_0^{II^*}n_{I}dV \\
    J_{b_1 b_1}& = \frac{\lambda_1 c}{2}\int \Psi_1^{I}\Psi_1^{I^*}n_{II}dV;\, 
    J_{b_2 b_2} = \frac{\lambda_1 c}{2}\int \Psi_1^{II}\Psi_1^{II^*}n_{I}dV.
\end{align}
Due to the symmetry, the expressions in each line of Eq.~(\ref{eq:self_freq}) are equal. Applying the notations from the main text, we obtain $\omega_1 = J_{a_1 a_1}$ and $\omega_2= J_{b_1 b_1}$.

The offdiagonal coefficients in system~(\ref{eq:sys_1}) depict the mode interaction. Let us consider the interaction between the fundamental and higher order modes of one resonator. The corresponding coefficients are expressed as
\begin{align}\label{eq:intra_1}
    J_{a_1 b_1} &= \frac{\lambda_1^2 c}{2 \lambda_0}\int \Psi_1^{I}\Psi_0^{I^*}n_{II}dV e^{-i(\omega_1 - \omega_0)t},\\
    J_{a_2 b_2} &= \frac{\lambda_1^2 c}{2 \lambda_0}\int \Psi_1^{II}\Psi_0^{II^*}n_{I}dV e^{-i(\omega_1 - \omega_0)t},\\
    J_{b_1 a_1} &= \frac{\lambda_0^2 c}{2 \lambda_1}\int \Psi_0^{I}\Psi_1^{I^*}n_{II}dV e^{-i(\omega_0 - \omega_1)t},\\
    J_{b_2 a_2} &= \frac{\lambda_0^2 c}{2 \lambda_1}\int \Psi_0^{II}\Psi_1^{II^*}n_{I}dV e^{-i(\omega_0 - \omega_1)t}.
\end{align}
As one can see, the interaction efficiency is enhanced at the points of degeneracy, where the eigenfrequencies coincide. These points correspond to the exact positions of the mode crossings. In the main text we consider this particular example, thus $J_{ab}^{\mathrm{int}} = J_{a_1 b_1}|_{\omega_0 = \omega_1}= J_{a_2 b_2}|_{\omega_0 = \omega_1}$ and $J_{ba}^{\mathrm{int}} = J_{b_1 a_1}|_{\omega_0 = \omega_1}=J_{b_2 a_2}|_{\omega_0 = \omega_1}$, where we assumed the coupling purely real for simplicity.

The coupling coefficients between the fundamental modes of both resonators take form:
\begin{align}
    J_{a_1 a_2} &= \frac{\lambda_0 c}{2}\int \Psi_0^{II}\Psi_0^{I^*}n_{I}dV,\\
    J_{a_2 a_1} &= \frac{\lambda_0 c}{2}\int \Psi_0^{I}\Psi_0^{II^*}n_{II}dV,
\end{align}
and they are equal due to the symmetry. The corresponding coefficient in the main text $J_{aa}^{\mathrm{ext}} = J_{a_1 a_2}J_{a_2 a_1}$ . In  the similar way, we express coupling between higher order modes
\begin{align}
    J_{b_1 b_2} &= \frac{\lambda_1 c}{2}\int \Psi_1^{II}\Psi_1^{I^*}n_{I}dV,\\
    J_{b_2 b_1} &= \frac{\lambda_1 c}{2}\int \Psi_1^{I}\Psi_1^{II^*}n_{II}dV,
\end{align}
with $J_{bb}^{\mathrm{ext}} = J_{b_1 b_2}= J_{b_2 b_1}$. 

The coefficients governing interactions between fundamental and higher order modes of distinct resonators are placed on the side diagonal of the system~(\ref{eq:sys_1}), and their expressions are by
\begin{align}\label{eq:inter_1}
    J_{a_1 b_2} &= \frac{\lambda_1^2 c}{2 \lambda_0}\int \Psi_1^{II}\Psi_0^{I^*}n_{I}dV e^{-i(\omega_1 - \omega_0)t},\\
    J_{a_2 b_1} &= \frac{\lambda_1^2 c}{2 \lambda_0}\int \Psi_1^{I}\Psi_0^{II^*}n_{II}dV e^{-i(\omega_1 - \omega_0)t},\\
    J_{b_2 a_1} &= \frac{\lambda_0^2 c}{2 \lambda_1}\int \Psi_0^{I}\Psi_1^{II^*}n_{II}dV e^{-i(\omega_0 - \omega_1)t},\\
    J_{b_1 a_2} &= \frac{\lambda_0^2 c}{2 \lambda_1}\int \Psi_0^{II}\Psi_1^{I^*}n_{I}dV e^{-i(\omega_0 - \omega_1)t}.
\end{align}
As one can see, the interaction increases at degenerate frequencies, and then $J_{ab}^{ext} = J_{a_1 b_2}|_{\omega_1 = \omega_2} = J_{a_2 b_1}|_{\omega_1 = \omega_2}$, $J_{ba}^{ext} = J_{b_2 a_1}|_{\omega_1 = \omega_2}= J_{b_1 a_2}|_{\omega_1 = \omega_2}$. It is important to note, that the intraresonator interaction originates from the mode overlap in the area where both modes decay exponentially (for example see Eq.~(\ref{eq:intra_1})), when the interesonator coupling originates from the area where one is localized and second one is evanescent (e.g. Eq.~(\ref{eq:inter_1})). However, it is hard to estimate the ratio between these coefficients because it also depends on the integral along azimuth coordinate. In order to obtain this ratio, we provide FDTD simulations, which are presented in the next sections.

\section{Matrix model}

Generalized to the case of N coupled resonators, the coupling matrix of the size 2Nx2N can be represented as follows:

\begin{equation}
i\frac{dU}{dt}=
  -  \begin{pmatrix}
-\omega_1 &  J_{aa}^{ext} &  &  &  J_{ab}^{int}& J_{ab}^{ext} &  & \\ 
 J_{aa}^{ext}& -\omega_1 & \ddots &  & J_{ab}^{ext} & J_{ab}^{int} & \ddots & \\ 
 & \ddots & \ddots &  J_{aa}^{ext}&  & \ddots & \ddots &  J_{ab}^{ext}\\ 
 &  &  J_{aa}^{ext}& -\omega_1 &  &  &  J_{ab}^{ext} &  J_{ab}^{int}\\ 
J_{ba}^{int} & J_{ba}^{ext} &  &  & -\omega_2 &  J_{bb}^{ext}&  & \\ 
J_{ba}^{ext} & J_{ba}^{int} & \ddots &  & J_{bb}^{ext}&-\omega_2 & \ddots  & \\ 
 & \ddots & \ddots &  J_{ba}^{ext}  &  & \ddots & \ddots & J_{bb}^{ext}\\ 
 &  & J_{ba}^{ext}  & J_{ba}^{int}  &  &  & J_{bb}^{ext}& -\omega_2
\end{pmatrix}
U,
\end{equation}
where $U = (a_1,...,a_N,b_1, ...,b_N)^\intercal$.
The coupling matrix is comprised of four blocks of symmetric tridiagonal matrices, implying that empty spaces are zeros. For the calculations presented in the article we suppose that $J_{ab}^{ext,int}$ and $J_{ab}^{ext,int}$ are equal due to the apparent summery, therefore second and third blocks of the coupling matrix are identical. Blocks one and four are also set to be identical since the difference between $J_{aa}^{\mathrm{ext}}$ and  $J_{bb}^{\mathrm{ext}}$ will lead to a simple shift along the direction of the higher-order mode and the mode interaction therefore has to be examined at $\omega_1 = \omega_2 = 0$.

\begin{figure*}
    \centering
    \includegraphics[width=0.9\textwidth]{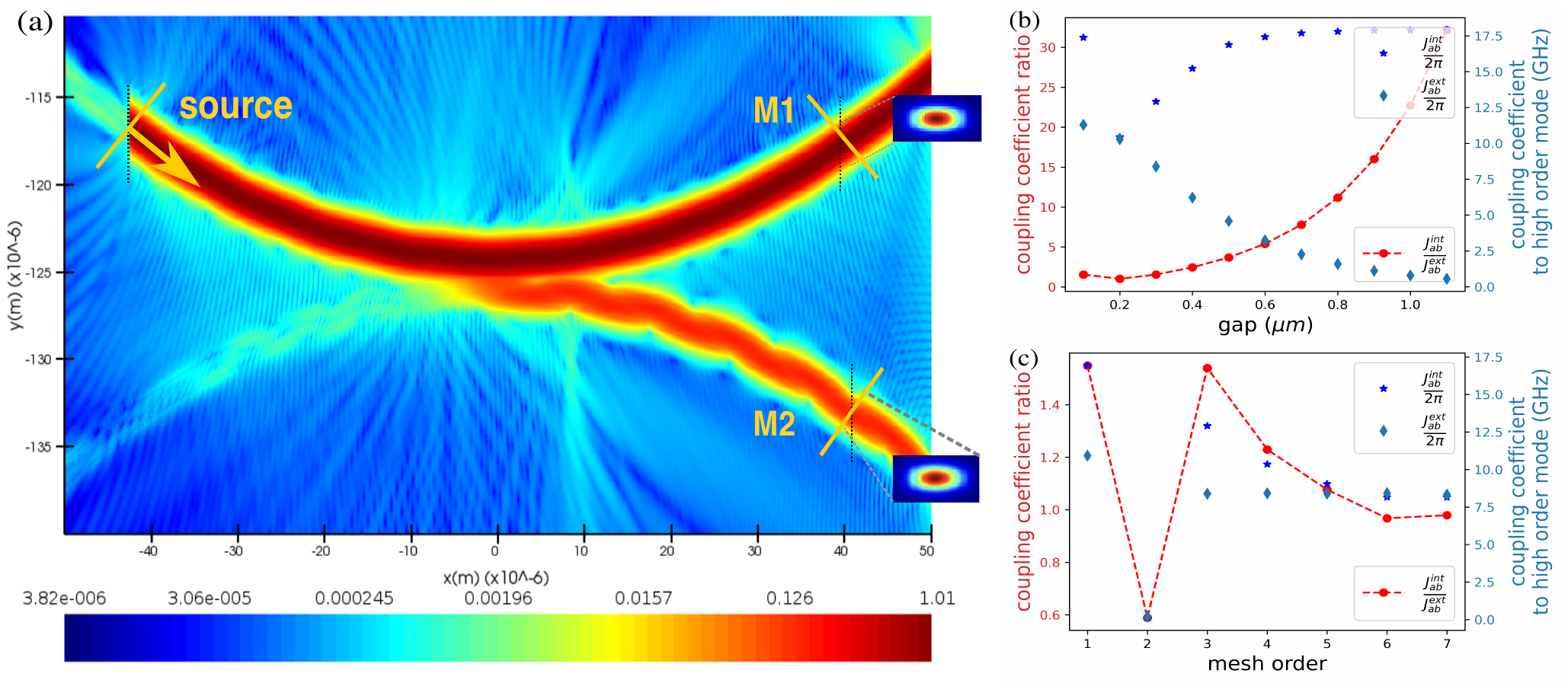}
    \caption{FDTD simulations of the coupling section. (a) Normalized electric field distribution shown in log scale, the colour bar measures the field power. The mode source is launched normal to the waveguide propagation direction with unity power. Insets present spatial field distributions recorded by monitor M1 and M2. The gap distance is set as 0.3 $\mu m$. (b) Dependence of the coupling coefficients to higher-order modes on gap distance, at mesh 3. (c) Convergence of the coupling coefficients to higher-order modes with mesh order, as illustrated by the red dashed line, the coupling coefficient ratio converges closely to unity.}
    \label{SI_fig:1}
\end{figure*}

\section{FDTD simulations of the coupling section}
In order to confirm the coupling coefficient ratio expected from the analytical model of four-mode interaction, we provide FDTD simulations of the coupling section of the photonic dimer. We constructed a model of a dimer device comprised of two 200 GHz ring resonators. The silicon nitride (Si$_{3}$N$_{4}$) resonator core is fully cladded with Silicon dioxide (SiO$_{2}$). Both resonators are 1.5 $\mu$m wide and 0.82 $\mu$m high, with sidewall angle $\alpha = 90^{\circ}$, as used in the experiments. The mode source was configured to inject at an angle of $20^{\circ}$ with unity power, as shown in Fig.~\ref{SI_fig:1}a, and to excite only the fundamental mode of the ring. In this way, we shrank the simulation region to 100 $\times$ 30 $\times$ 8 $\mu m^3$ and the simulation time to 900 fs, which is sufficient to capture correctly the coupling to higher-order modes with much less processing time. The boundary of the simulation region is fixed with a perfectly matched layer (PML) condition to absorb the incident light and therefore to prevent backreflection. The light field then propagated in the full simulation region until a stationary state is reached. Monitors M0, M1 and M2 recorded the spatial distributions of the mode source, the transmitted field and the coupled field respectively. In addition, two mode expansion monitors were placed in the same plane as M1 and M2 to calculate the power of selected resonator eigenmode (TE$_{10}$). All powers are normalized as they derived from the resonator fundamental mode that is launched with unity power. The coupling coefficients, J$_{ab}^{int}$ and J$_{ab}^{ext}$, are estimated using a simplified coupled mode equation, by J$_{ab}^{i}$ = $D_1 \times \arccos(\sqrt{1-P_{ab}^{i}})$, where $D_1/2\pi =FSR$.

Numerical simulations reveal that increasing the gap distance between two resonators, J$_{ab}^{ext}$  rapidly (eventually exponentially) decays, while J$_{ab}^{int}$ remains constant at 18 GHz except the region 0.2-0.4 $\mu$m where it demonstrates lower values at mesh order 3 as shown in Fig.~\ref{SI_fig:1}. Careful analysis of the ratio J$_{ab}^{int}/$J$_{ab}^{ext}$ convergence with increasing mesh order (decreasing the simulation net size) suggests that the ratio converges to unity. The simulations with mesh order 3 gives a considerable error of $\approx 35 \%$.

The coupling of the ring resonator to the bus and drop waveguides was also simulated at 0.3 $\mu$m, rendering a converged result of 0.75 GHz, which contributes to the coefficient $J_{ab}^{\mathrm{int}}$.